%% file: main.tex
\documentclass[acmsmall,manuscript,screen]{acmart}

\usepackage{tabularx}
\usepackage{enumitem}
\usepackage{csquotes}
\usepackage{pdflscape}
\usepackage[most]{tcolorbox}
\usepackage{needspace}
\usepackage{todonotes}
\usepackage{longtable, booktabs}

\newcolumntype{Y}{>{\centering\arraybackslash}X}

\tcbset{
  boxrule=0.5pt,
  colframe=black!35,
  colback=black!3,
  coltext=black,
  coltitle=black,
  colbacktitle=black!3,
  sharp corners,
  before skip=\baselineskip,
  after skip=\baselineskip,
}

\newtcolorbox{answerbox}[2][]{%
  enhanced, breakable,
  title={#2},
  attach title to upper,
  minipage boxed title,
  after title={\enspace},
  before skip=10pt,
  fonttitle=\bfseries,
  #1
}

\setlist[description]{
  style=standard,
  labelwidth=2.5cm,
  labelsep=1em,
  leftmargin=!,
  align=left
}

\newenvironment{dimensiondef}[1][]{
  \par\needspace{1\baselineskip}
  \noindent
  {\bfseries #1}\par\penalty10000\smallskip
  \noindent{\small\MakeUppercase{Definition}}
  \noindent
}{\par\smallskip}

\newenvironment{examples}{
  \par\needspace{3\baselineskip}
  \noindent{\small\MakeUppercase{Examples}}
  \nopagebreak
  \vspace{-0.25\baselineskip}
  \noindent
  \begin{list}{}{
    \setlength{\leftmargin}{0.5em}
    \setlength{\labelwidth}{0pt}
    \setlength{\labelsep}{0.5em}
    \setlength{\itemindent}{0pt}
    \setlength{\listparindent}{0pt}
    \setlength{\rightmargin}{0pt}
    
  }
}{
  \end{list}\par\smallskip
}

\newenvironment{origin}{
  \par\needspace{3\baselineskip}
  \noindent{\small\MakeUppercase{Origin}}
  \nopagebreak
  \vspace{-0.25\baselineskip}
  \noindent
  \begin{list}{}{
    \setlength{\leftmargin}{0.5em}
    \setlength{\labelwidth}{0pt}
    \setlength{\labelsep}{0.5em}
    \setlength{\itemindent}{0pt}
    \setlength{\listparindent}{0pt}
    \setlength{\rightmargin}{0pt}
    
  }
}{
  \end{list}\par\smallskip
}

\AtBeginDocument{%
  }

\setcopyright{cc}
\setcctype{by-sa}
\copyrightyear{2025}
\acmYear{2025}
\acmDOI{XXXXXXX.XXXXXXX}

\begin{document}

\title{Quality of Descriptive Information on Cultural Heritage Objects: Definition and Empirical Evaluation}

\author{Markus Matoni}
\orcid{https://orcid.org/0000-0003-4389-5871}
\affiliation{
  \institution{Gesellschaft für wissenschaftliche Datenverarbeitung mbH Göttingen}
  \city{Göttingen}
  \country{Germany}
}
\email{markus.matoni@gwdg.de}

\author{Arno Kesper}
\orcid{https://orcid.org/0000-0002-5042-1087}
\affiliation{
  \institution{Philipps-Universität Marburg}
  \city{Marburg}
  \country{Germany}
}
\email{arno.kesper@uni-marburg.de}

\author{Gabriele Taentzer}
\orcid{https://orcid.org/0000-0002-3975-5238}
\affiliation{
  \institution{Philipps-Universität Marburg}
  \city{Marburg}
  \country{Germany}
}
\email{taentzer@uni-marburg.de}

\begin{abstract}
  \input{0_abstract}
\end{abstract}

\begin{CCSXML}
  <ccs2012>
  <concept>
  <concept_id>10002951.10003227.10003392</concept_id>
  <concept_desc>Information systems~Digital libraries and archives</concept_desc>
  <concept_significance>500</concept_significance>
  </concept>
  <concept>
  <concept_id>10002951.10002952.10003219.10003218</concept_id>
  <concept_desc>Information systems~Data cleaning</concept_desc>
  <concept_significance>500</concept_significance>
  </concept>
  <concept>
  <concept_id>10010405.10010469</concept_id>
  <concept_desc>Applied computing~Arts and humanities</concept_desc>
  <concept_significance>300</concept_significance>
  </concept>
  <concept>
  <concept_id>10002944.10011123.10010912</concept_id>
  <concept_desc>General and reference~Empirical studies</concept_desc>
  <concept_significance>100</concept_significance>
  </concept>
  <concept>
  <concept_id>10003456.10003457.10003490.10003507.10003510</concept_id>
  <concept_desc>Social and professional topics~Quality assurance</concept_desc>
  <concept_significance>100</concept_significance>
  </concept>
  </ccs2012>
\end{CCSXML}

\ccsdesc[500]{Information systems~Digital libraries and archives}
\ccsdesc[500]{Information systems~Data cleaning}
\ccsdesc[300]{Applied computing~Arts and humanities}
\ccsdesc[100]{General and reference~Empirical studies}
\ccsdesc[100]{Social and professional topics~Quality assurance}

\ccsdesc[500]{Information systems~Data structures}

\keywords{Data Quality, Quality Dimensions, Cultural Heritage, Quality Assurance, Data Cleaning, Uncertainty}

% \received{1 November 2025}

\maketitle

\input{1_introduction}
\input{2_0_basic_terms}

\input{3_related_work}
\input{4_0_data_quality_dimensions}

\input{5_0_empirical_evaluation}
\input{6_conclusion}

\bibliographystyle{ACM-Reference-Format}
\bibliography{references}

\input{appendix}

\end{document}

%% file: 0_abstract.tex
Effective data processing depends on the quality of the underlying data.
However, quality issues such as inconsistencies and uncertainties, can significantly impede the processing and subsequent use of data.
Despite the centrality of data quality to a wide range of computational tasks, there is currently no broadly accepted, domain-independent consensus on the definition of data quality.
Existing frameworks primarily define data quality in ways that are tailored to specific domains, data types, or contexts of use.
Although quality assessment frameworks exist for specific domains, such as electronic health record data and linked data, corresponding approaches for descriptive information about cultural heritage objects remain underdeveloped.
Moreover, existing quality definitions are often theoretical in nature and lack empirical validation based on real-world data problems.
In this paper, we address these limitations by first defining a set of quality dimensions specifically designed to capture the characteristics of descriptive information about cultural heritage objects.
Our definition is based on an in-depth analysis of existing dimensions and is illustrated through domain-specific examples.
We then evaluate the practical applicability of our proposed quality definition using a curated set of real-world data quality problems from the cultural heritage domain.
This empirical evaluation substantiates our definition of data quality, resulting in a comprehensive definition of data quality in this domain.

%% file: 1_introduction.tex
\section{Introduction}\label{sec:introduction}
In an increasingly data-driven world, data quality (DQ) directly influences the reliability and effectiveness of decision-making, research, and operational processes.
Poor DQ can compromise analytical outcomes and lead to erroneous conclusions.
Therefore, ensuring DQ has become a central concern across numerous domains \citep{batini2009}.
A key challenge in assessing DQ lies not only in detecting quality issues but also in \emph{making the quality of data explicit} \citep{wang1996,pipino2002}.
Achieving this requires defining DQ for a specific domain and context of use to ensure that DQ can be evaluated in a systematic and reproducible manner.

The concept of \emph{quality dimensions} has emerged as an essential concept for defining DQ.
However, dimension-based DQ definitions can vary widely, substantially in both the selection of dimensions and how they are defined.
Foundational works, such as those by \citet{wang1996}, and the ISO~25012~standard~\citep{iso:25012}, define DQ as a multifaceted concept consisting of a set of quality dimensions, each capturing a distinct aspect of DQ.
Dimensions such as accuracy, completeness, and timeliness represent examples for quality aspects or criteria and provide a structured lens through which data can be evaluated \citep{wang1996,batini2009,iso:25012}.
These dimensions can be adapted to guide quality assessment frameworks, data governance strategies, and quality improvement methodologies.
Despite certain recurring elements across the definitions presented in the literature, \emph{there is no consensus on a single, generally accepted DQ definition} \citep{matoni2025}.

While the above-mentioned works present domain-independent DQ definitions, many other DQ definitions are tailored to a specific domain or data types.
Examples include the DQ definition for healthcare by \citet{chiasera2011} and the quality definition for linked open data (LOD) by \citet{zaveri2015}.
It is widely acknowledged that DQ is inherently context-dependent \citep{wang1996,kahn2016,ehrlinger2022} and should therefore be tailored to a specific domain or data type \citep{eppler2000,caro2008,sadiq2018}.
The relevance, interpretation, and operationalization of individual quality dimensions can vary significantly across domains, depending on the respective goals, characteristics, and constraints \citep{caro2008,sadiq2018}.
Consequently, each domain bears the responsibility of defining its own quality dimensions.
However, \emph{within the cultural heritage domain, a comprehensive and widely accepted definition of DQ specifically for descriptive information about cultural heritage objects remains lacking}.
This absence hinders efforts to develop effective quality assurance techniques and tools, as well as the reusability and interoperability of data \citep{dallas2015,zeng2019}.

While several domains have developed tailored DQ definitions, the cultural heritage community has not yet undertaken a systematic review to determine whether existing DQ definitions can be reused or adapted.
Without such a review, it remains unclear whether foundations for defining DQ in the cultural heritage domain should be derived from existing works or developed anew.

In the cultural heritage domain, descriptive information is used to describe artworks, monuments, and other historical artifacts.
This information typically includes titles, creators, materials, techniques, and contextual narratives \citep{ranjgar2019,beaudoin2012,salse-rovira2024}.
It constitutes a fundamental resource for research, preservation, and public dissemination.
In this domain, this type of information is also referred to as (descriptive) metadata, which is distinct from purely administrative and technical metadata \citep{riley2017}.
Referred to as  metadata in this domain, such information is distinguished from purely administrative and technical metadata.
For example, the creator of a painting is considered descriptive metadata, whereas file formats or modification timestamps represent technical metadata.
Cultural heritage institutions rely on descriptive metadata to provide content- and domain-specific documentation that supports scholarly and curatorial practice \citep{lorenzini2021,baca2008}.
Despite its centrality, descriptive metadata exhibits distinctive structural and semantic characteristics that pose challenges to quality assessment and interoperability.

\emph{Descriptive information about cultural heritage objects is typically semi-structured}, and only in some cases fully structured \citep{zeng2019,hyvonen2012}.
Semi-structured data may be associated with an underlying schema or data model.
Although institutions frequently adopt established schemas, such as Dublin Core \citep{dublincore-spec}, CIDOC CRM \citep{cidoc-crm}, or LIDO \citep{lido-spec}, their implementations often vary, resulting in data that is only partially constrained by formal structure \citep{felicetti2010,huggett2012}.
As \citet{dallas2015} notes, cultural heritage data is inherently heterogeneous, interpretive, and shaped by specific curatorial, disciplinary, and epistemological practices.
The semi-structured nature of data complicates efforts to validate data, enforce consistency, and support automated reasoning, thereby underscoring the need for domain-appropriate approaches to DQ assessment.

Another important characteristic of descriptive information about cultural heritage objects is its emphasis on semantic linking, both within individual collections and across institutional boundaries.
Cultural heritage institutions increasingly seek to publish and reuse data as LOD, a process that entails aligning locally maintained data with external vocabularies and authority files, such as the Getty Vocabularies~\citep{getty-vocabularies}, Wikidata~\citep{wikidata}, and the GND~\citep{DNB_GND} \citep{candela2019,baca2015}.
While such linking enables semantic enrichment, enhanced discovery, and improved data reuse, it also introduces additional complexity \citep{dressen2025,candela2019}.
Maintaining the consistency and meaningfulness of connections across heterogeneous datasets requires careful attention to potential quality issues.
Understanding the nature and quality of descriptive data in the cultural heritage context is therefore essential for ensuring its reliability and enabling its broader use in digital humanities research, knowledge graphs processing, and cultural analytics.
Below, we refer to descriptive information about cultural heritage objects as \emph{object description data}.

Defining DQ for object description data requires consideration of dimensions that go beyond core quality aspects such as accuracy or completeness, and that reflect the particular ways object description data is created, curated, and reused.
Certain aspects become especially critical, including provenance, which has repeatedly been identified as essential in the cultural heritage context \citep{mckenna2022,massari2025,alkemade2025}.
Provenance captures the sources needed to verify and evaluate the trustworthiness and interpretability of information about cultural heritage objects \citep{moreau2013}.
In contexts where records combine curatorial expertise with external authority sources, understanding how and why a particular date, attribution, or contextual note was introduced is vital for establishing scholarly reliability \citep{doerr2011}.
Closely related is the notion of causal traceability, the capability to reconstruct how information has evolved over time (for example, through successive curatorial decisions).
Such traceability supports the resolution of conflicting information and facilitates the integration of heterogeneous datasets \citep{barabucci2022,daquino2022}.
It further enables advanced reuse scenarios, including LOD publication and digital humanities analyses \citep{massari2025,alkemade2025,camara2023}.
Yet, existing DQ definitions rarely incorporate these aspects in ways that adequately account for the interpretive, heterogeneous, and semi-structured nature of object description data \citep{hannah2014}.
This gap underscores the \emph{need for a DQ definition for object description data that explicitly integrates these domain-specific aspects}.

This paper aims to address these challenges by developing a domain-specific, empirically grounded definition of DQ for descriptive information about cultural heritage objects.
We begin by examining how existing definitions and standards may apply within this context.
We then develop a comprehensive DQ definition tailored to the distinctive characteristics of object description data.
Finally, we assess the practical utility of this definition by evaluating it against real-world quality issues.
To this end, we investigate the following research questions (RQs):
\begin{enumerate}[label= \textbf{RQ\arabic*}, leftmargin=*, ref=RQ\arabic*]
      \item\label{rq:related-work} What existing quality definitions are applicable to descriptive information about cultural heritage objects?
      \item\label{rq:definition} How should the quality of descriptive information about cultural heritage objects be defined?
      \item\label{rq:evaluation} Is the proposed data quality definition for descriptive information of cultural heritage objects appropriate?
\end{enumerate}

To address these research questions, this paper makes the following contributions:
\begin{enumerate}[label= (\arabic*), leftmargin=*]
      \item \emph{Mapping Existing Literature}:
            Drawing on the body of publications identified in \citet{matoni2025}, we determine which existing DQ definitions are appropriate for descriptive information about cultural heritage objects.
      \item \emph{Quality Definition}: Using the selected publications, we define a set of quality dimensions and subdimensions tailored to descriptive information about cultural heritage objects.
            Our approach results in a domain-specific definition of DQ that specifies the origin of each dimension and illustrates each with realistic examples from the cultural heritage context.
      \item \emph{Empirical Evaluation}: We evaluate the proposed definition using a set of empirically collected quality problems from the cultural heritage domain.
            These problems include heterogeneous structural representations arising from the integration of multiple data sources, implicitly encoded uncertainties, and inconsistent use of vocabularies.
            We examine whether all collected quality problems can be classified using the identified quality dimensions, whether all these dimensions are necessary for classification, and whether the defined quality dimensions are mutual orthogonal.
\end{enumerate}

We structure this work into the following sections: Section~\ref{sec:basic_terms} introduces the technical concepts and terminology used throughout this paper.
Section~\ref{sec:related_work} reviews existing DQ definitions, evaluating their relevance to descriptive information about cultural heritage objects.
Section~\ref{sec:data_quality_dimensions} presents our own dimension-based definition tailored to this domain.
Section~\ref{sec:empirical_evaluation} then assesses the definition against a catalog of empirically observed quality problems in cultural heritage data.
Finally, Section~\ref{sec:conclusion} summarizes the findings and outlines implications for future research.

%% file: 2_0_basic_terms.tex
\section{Basic Terms}\label{sec:basic_terms}
This section introduces the terminology relevant to data quality within the cultural heritage domain.
By providing clear definitions of key concepts, including data, quality dimensions, and linked data, we establish a shared vocabulary that underpins the discussions and analyses presented in the subsequent sections.

\subsection{Data and Information}
\input{2_1_data_information.tex}

\subsection{Data Quality and Quality Dimension}
\input{2_2_data_quality.tex}

\subsection{Subdimension}
\input{2_3_sub_dimensions.tex}

\subsection{Domain of Interest and Context of Use}
\input{2_4_domain_context.tex}

\subsection{Dataset, Data Element, Data Property, Data Value and Link}
\input{2_5_dataset_element_property_value_link.tex}

\subsection{Semi-structured Data and Data Model}
\input{2_6_semi_structured_data_model.tex}

\subsection{Linked (Open) Data}
\input{2_7_linked_open_data.tex}

\subsection{Descriptive Information about Cultural Heritage Objects}
\input{2_8_descriptive_information.tex}

%% file: 2_1_data_information.tex
\emph{Data} is a multifaceted concept with varying interpretations across domains.
The ISO~25012~standard~\citep{iso:25012} defines data broadly as a \enquote{reinterpretable representation of information in a formalized manner suitable for communication, interpretation, or processing}.
In turn, \emph{information} is defined as \enquote{knowledge concerning objects, such as facts, events, things, processes, or ideas, including concepts, that within a certain context have a particular meaning} \citep{iso:25012}.
Because the selection and representation of information in data require deliberate choices,
data can be understood as a model of the world that is of interest to a given context or purpose \citep{sebastian-coleman2022}.

%% file: 2_2_data_quality.tex
The definition of \emph{data quality} (DQ) has been a significant topic of research for decades \citep{wang1996,lacagnina2023}, giving rise to numerous DQ definitions that vary considerably in scope and focus.
Some foundational works, such as \citet{wang1996} and the ISO~25012 standard~\citep{iso:25012}, define DQ as a multifaceted concept composed of a set of quality dimensions, with each dimension capturing a distinct aspect of DQ.
While these definitions of DQ address data in general, other definitions focus on specific formats, such as linked data (e.g.,~\citet{zaveri2015}), or to specific domains, such as healthcare (e.g.,~\citet{chiasera2011}).
Quality dimensions, such as accuracy, completeness, and timeliness, collectively provide a conceptual framework for understanding DQ.
However, existing DQ definitions differ substantially both in the dimensions they include and in how these dimensions are defined.
To organize this diversity, quality dimensions are often grouped into categories, such as intrinsic, contextual, and representational DQ (cf.~\citet{wang1996}).
In this context, a \emph{dimension} denotes a fundamental, qualitative characteristic that represents a specific aspect of a data set's fitness for its intended purpose.
Dimensions are inherently qualitative and serve as the conceptual basis for assessing DQ at a conceptual level, while metrics provide measurable, quantitative expressions that operationalize these dimensions, forming the next step in a structured quality assurance process.

%% file: 2_3_sub_dimensions.tex
A \emph{subdimension} represents a more specific, lower-level aspect of a broader dimension.
In hierarchical models, high-level dimensions are decomposed into subdimensions to capture the finer-grained components of a concept and allow for more precise assessment and targeted measurement.
In practice, multifaceted dimensions are often subdivided,
with each subdimension defined individually to provide a detailed understanding of data quality.
Formally, a subdimension can be considered a subset of the parent dimension's structure \citep{fries1991,mansmann2007}.
While subdimensions are treated as conceptually independent for analytical purposes, they are explicitly associated with a higher-level dimension to maintain the hierarchical organization of the quality framework.

%% file: 2_4_domain_context.tex
In general, data represents information about a specific domain of interest.
A \emph{domain of interest} refers to a specific area or subject of the real world on which the data is focused.
In the context of knowledge organizations, a domain constitutes a body of knowledge shared by a community, shaped by common concepts and practices \citep{hjorland2017}.
Domain can vary in scope: they may be broad, such as cultural heritage, encompassing data curation and digital collections, or more narrowly defined, such as architecture within cultural heritage, which focuses on managing descriptive information about the design, construction, and preservation of buildings \citep{ziku2020}.
While the domain establishes the scope of data, it does not by itself determine its use.
Data selection and representation are also influenced by the \emph{context of use}, including the questions being addressed, the tasks required to process the data, and the tools employed to perform these tasks.
Typical data processing tasks include searching and analyzing data, data mining, data profiling, data cleansing, and data integration.
In this paper, we consider cultural heritage as an example of a domain of interest, which can be further refined to subdomains such as architecture.
The primary context of use in the cultural heritage involves information retrieval to support addressing research questions.
Key concepts of this domain include objects such as paintings, persons associated with an object, locations, and images.
Domain experts play a central role in determining both the domain and its boundaries, as well as the relevant contexts of use.
In cultural heritage, these experts typically include librarians, curators, custodians, and researchers.
For data to be \emph{fit for use}~\citep{wang1996}, it is essential to examine the specific DQ requirements associated with each context of use, ensuring that the data meets the needs of its intended applications.

%% file: 2_5_dataset_element_property_value_link.tex
We adopt the following terminology to refer to descriptive information about cultural heritage objects:
\emph{Data} may refer to a single data value, a collection of related values, or an entire dataset of interrelated data elements.
A dataset consists of a set of data elements.
A \emph{data element} --- also called data item, entity instance, record, row, or node --- has a set of properties and a set of links.
A \emph{property} --- also called a property node, field, attribute, or column --- holds a \emph{data value}, which may be a string, a number, a Boolean, or of another primitive type.
A \emph{link} --- also called a column, arc, or edge --- is a reference from one data element to another, either within the same dataset or across different datasets; it may also refer to the element itself.
In the context of linked data, an \emph{interlink} is a single, explicit connection, such as a URL or RDF triple, that links an entity in one dataset to a related or corresponding entity in another.
The term \emph{interlink} emphasizes the cross-dataset nature of the connection, and aligns with current usage in the LOD literature of \citet{zaveri2015} and \citet{mckenna2022}.

%% file: 2_6_semi_structured_data_model.tex
Descriptive information about cultural heritage objects is usually semi-structured and may or may not conform to an explicit data model, making it necessary to clarify these terms.
Data can be distinguished based on its form:
According to \citet{abiteboul1997}, \emph{semi-structured data} \enquote{resides in different forms, ranging from unstructured data in file systems to highly structured in relational database systems}, and \enquote{is (from a particular viewpoint) neither raw data nor strictly typed, i.e., not table-oriented as in a relational model or sorted-graph as in object databases}.
Structured data is always accompanied by a data model that explicitly defines its structure.
In the relational model, structured data is represented as interrelated data entities with named attributes, each holding a data value.
In contrast, unstructured data, such as text, images, and video, lacks an explicit internal or external structure.
Semi-structured data is represented using a data description language such as XML that pre-structures the data.
As noted by \citet{ambika2020} semi-structured data is \enquote{a combination of structured and unstructured data and shares characteristics of both}.
When XML data is not associated with an XML schema, there is no external representation of its structure, a property referred to as self-describing.
The languages, techniques, and tools used to process data depend on its structural characteristics.
Structured data always has an underlying data model, whereas semi-structured data may or may not.
As \citet{mohagheghi2009} observe, \enquote{models are representations of a (software) system at an abstract level} and are \enquote{developed using a modelling language}.
Various types of models exist to represent different aspects of systems, with data models specifically representing data structures.
A data model is a graphical and/or textual representation that specifies the properties, structure, and interrelationships of data within a given domain.
In the context of relational data or XML data, a data model is often referred to as a schema.

%% file: 2_7_linked_open_data.tex
As cultural heritage institutions increasingly publish and connect their data, descriptive information about cultural heritage objects is adopting characteristics of \emph{Linked Data}~(LD) and \emph{Linked Open Data}~(LOD), necessitating clarification of these concepts.
A common practice is interlinking, for example, in which collection records are linked to authority files, controlled vocabularies, or external databases, including the Virtual International Authority File (VIAF)~\citep{oclcinc.2025} and Wikidata~\citep{wikidata}.
The LD \enquote{principles promote publishing data and interlinking them in a machine-readable manner using Web standards}\citep{radulovic2017}.
The major principles of LD, formerly articulated by Tim Berners-Lee~\citep{berners-lee2009} in 2006, are:
\enquote{(1) Use URIs as names for things, (2) use HTTP URIs so that people can look up those names, (3) when someone looks up a URI, provide useful information, using the standards (RDF, SPARQL), and (4) include links to other URIs, so that they can discover more things}~\citep{berners-lee2009}.

LOD builds on these principles by emphasizing openness, ensuring that the data is publicly accessible \citep{paquet2020}.
Both LOD and LD facilitate the connection of heterogeneous data sources through explicit links, thereby integrating and reusing data across organizational boundaries.
This data integration is considered lightweight because it does not require transforming data into different structures or adapting it to a single ontology; rather, data can retain its syntactic and semantic heterogeneity while still being connected and interoperable.

%% file: 2_8_descriptive_information.tex
\begin{figure}[ht]
  \centering
  \begin{minipage}[t]{0.35\textwidth}
    \centering
    \includegraphics[width=\linewidth]{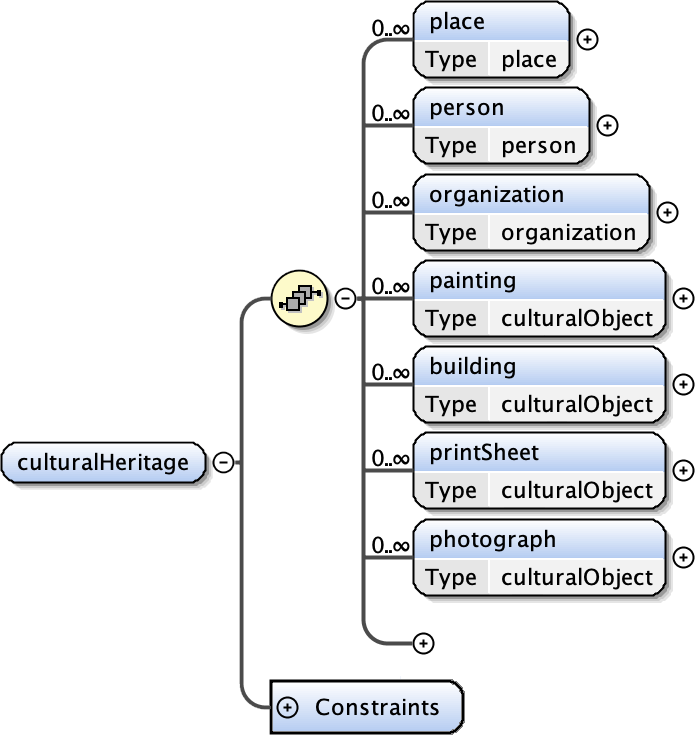}
  \end{minipage}
  \hfill
  \begin{minipage}[t]{0.63\textwidth}
    \centering
    \includegraphics[width=\linewidth]{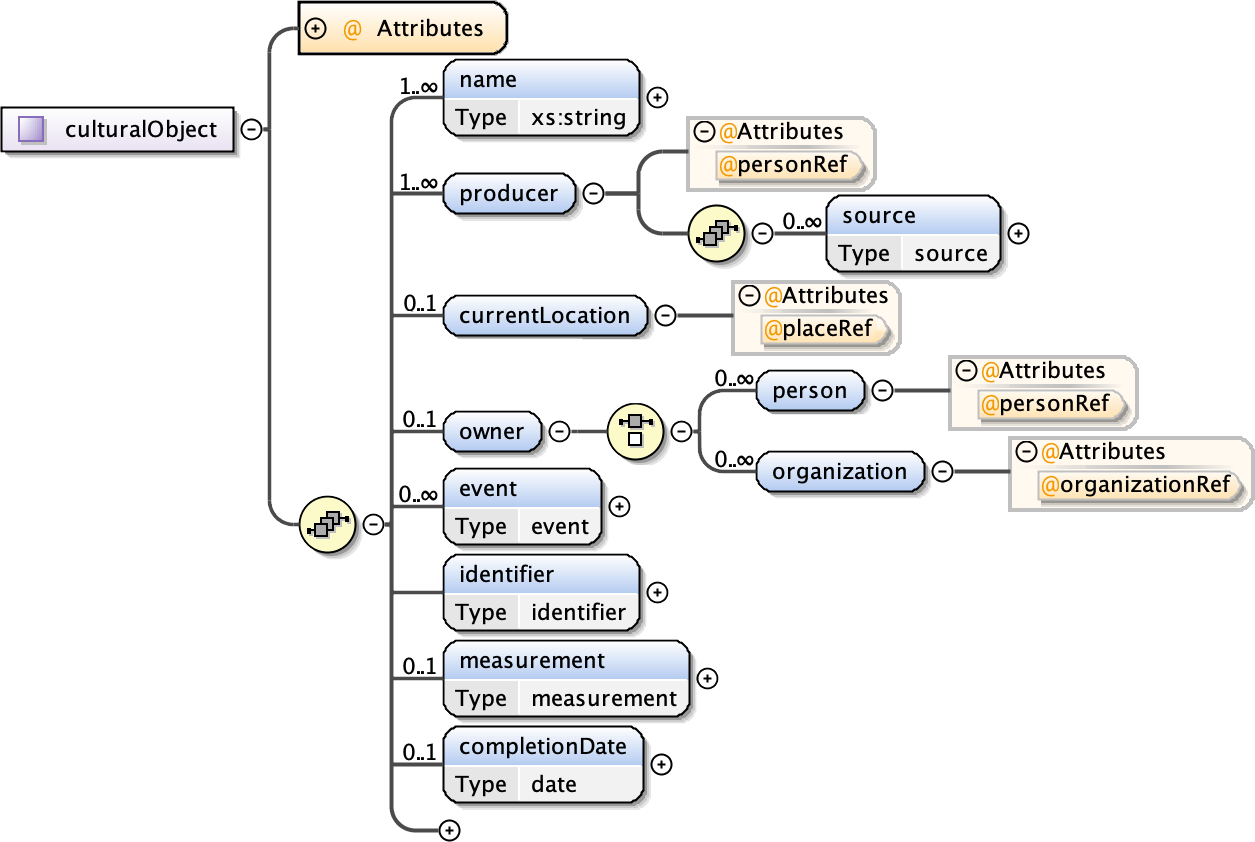}
  \end{minipage}
  \caption{
    This figure is one of two illustrations in this paper that show an example data model from the cultural heritage domain.
    It shows the elements \texttt{culturalHeritage} and \texttt{culturalObject}, both implemented as \texttt{complexType}.
  }\label{fig:demoModelCulturalObject}
  \Description{
    This figure is one of two illustrations in this paper that show an example data model from the cultural heritage domain.
    It shows the elements \texttt{culturalHeritage} and \texttt{culturalObject}, both implemented as \texttt{complexType}.
  }
\end{figure}

In Section~\ref{sec:introduction}, we defined the scope of \emph{descriptive information about cultural heritage objects}, which we also refer to as object description data.
To illustrate this type of data and highlight its similarities with semi-structured and LOD, we present a representative data model that demonstrates key terminology and aspects.
Figure~\ref{fig:demoModelCulturalObject} provides excerpts from a data model defined in the form of an XML schema.
This schema exemplifies how cultural objects can be recorded in a cultural heritage database.
The model element \texttt{CulturalHeritage} serves as a container for various types of cultural objects, including buildings, paintings, printed sheets, and photographs.
Each \texttt{CulturalObject} may include descriptive information such as a \texttt{name}, a \texttt{producer}, a \texttt{current location}, and associated \texttt{events}, thereby capturing essential metadata for research, curation, and data integration purposes.

%% file: 3_related_work.tex
\section{Related Work}\label{sec:related_work}
This section examines whether existing definitions of data quality (DQ) address the open problems outlined in Section~\ref{sec:introduction}.
To addresses~\ref{rq:related-work}~(\emph{What existing quality definitions are applicable to descriptive information about cultural heritage objects?}), we analyze how prior work engages with the following three key characteristics of descriptive information about cultural heritage objects:
\begin{enumerate}[label=(\roman*), leftmargin=*]
  \item \textbf{Semi-structuredness}:
        Semi-structured data combines constrained elements, such as XML elements controlled by XSD, Schematron rules that compensate for XSD limitations, and controlled vocabularies for materials or techniques, with unconstrained elements, including free-text descriptions, contextual narratives, and explicit representation of uncertainty.
        In addition, semi-structured data is frequently shaped by implicit, manually applied curatorial rules that vary across institutions and domains.
  \item \textbf{Link quality}:
        As cultural heritage data is increasingly published as Linked Open Data (LOD), we consider link quality becomes a critical concern.
        We consider link validity (i.e., resolvability and persistence of links), integrity (i.e., the technical and semantic reliability of the link target), and accuracy (i.e., the correct identification of the intended entity).
  \item \textbf{Domain-specific aspects}:
        Cultural heritage is shaped by aspects of research and preservation.
        These include the documentation of information provenance, the ability to trace changes across time and workflow steps, the trustworthiness of data sources, compliance with community and institutional documentation standards, and the explicit representation of uncertainty.
\end{enumerate}

A substantial body of literature defines DQ in terms of sets of quality dimensions that capture these divers quality aspects.
Some publications address DQ in general, while others are tailored to specific domains, such as electronic health record data (\citet{chiasera2011}), or to specific data formats, such as LOD (\citet{zaveri2015}), or specific quality aspects, such as ethics (\citet{firmani2020}).
The systematic survey presented in \citep{matoni2025} identifies and classifies peer-reviewed publications that propose original, dimension-based definitions of DQ.
We base our related work considerations on the outcome of this survey, which offers a structured and comprehensive foundational study grounded in a structured literature search.
This study identified publications that explicitly define quality dimensions, rather than merely providing metrics or assessment techniques.
A dimension expresses a prescriptive condition that characterizes what it means for data to be of high quality.
In contrast, a metric provides an operationalization of a quality definition, such as the proportion of recorded creators relative to the total number of known creators for a given object.

The comprehensive set of 35 publications identified in \citep{matoni2025} forms the basis for our analysis.
For the purpose of our study, however, we focus on a carefully selected subset of these works.
This subset includes only those quality definitions that formulate \emph{requirements} rather than merely listing \emph{attributes}.
Requirement-based definitions formulate quality in a prescriptive and declarative way, stating explicit goals that can be assessed.
For example, a requirement-based definition of completeness states that it \emph{refers to the degree to which a dataset contains all data elements of the domain of interest}.
In contrast, attribute-based definitions simply list desirable attributes, for example, describing completeness as \emph{sufficient}, \emph{comprehensive}, and \emph{includes all necessary values}.

In addition, we include only those quality definitions whose domain of interest is either generic or closely aligned with the context of descriptive information about cultural heritage objects.
This criterion ensures that the selected literature is both conceptually applicable and contextually relevant.
The resulting subset contains quality definitions most pertinent to our scope.
In the following, we discuss these definitions in detail and assess their adaptability to descriptive information about cultural heritage objects.

\emph{\citet{wand1996}.} They present a seminal definition of DQ that is primarily designed for structured data within information systems, such as relational databases.
While their definition focuses on intrinsic dimensions such as accuracy and completeness, it lacks dimensions essential for cultural heritage, including syntactic accuracy and conciseness.
Moreover, the definitions presuppose the existence of a well-defined conceptual schema with identifiable entities, attributes, and lawful states.
For example, accuracy is defined in terms of how well the system reflects a real-world state.
Context of use and compliance with community documentation standards are only implicitly addressed.
With regard to our open problems, \citet{wand1996} focus on the intrinsic dimensions of structured data rather than the mixed nature of semi-structuredness.
They also do not cover link quality and domain-specific aspects, such as provenance of information.

\emph{\citet{scannapieco2004}.} They focus on data from cooperative information systems and propose four quality dimensions.
Although their definitions of `consistency' and `accuracy' exhibit some awareness of semi-structuredness, the remaining dimensions are articulated at a high level, limiting their applicability to object description data.
For instance, `currency' is defined solely in terms of value changes, whereas updates in the cultural heritage are often event-driven (e.g., new scholarly attributions) and may affect elements and properties as well.
As with \citet{wand1996}, link quality is not addressed at all.
Consequently, the work focuses on a concise set of intrinsic dimensions, with limited attention to semi-structuredness, link quality, and specific cultural heritage requirements such as provenance and traceability.

\emph{ISO~25012~\citep{iso:25012}.} The work provides a general and comprehensive standard for DQ of structured data within computer systems.
It provides 15 DQ dimensions that jointly assess how intrinsic and system-context attributes contribute to the data's fitness for intended use.
However, the standard neither traces the conceptual origins of these dimensions nor links them to prior research.
Regarding our open problems, ISO~25012~\citep{iso:25012} does not address aspects of semi-structured data or link quality, such as validity and accuracy of interlinks.
Additionally, it offers limited guidance on assessing the trustworthiness of sources or corroborating statements in scholarly contexts.

\emph{\citet{zaveri2015}.} They introduce a quality framework for LOD, which incorporates linkage aspects more thoroughly than earlier definitions.
However, many of the defined dimensions are closely tied to the assumptions of the Resource Description Framework (RDF) serializations and web stack assumptions, limiting their direct applicability to object description data.
While the definitions’ provenance is based on a survey of existing work, the resulting synthesis is not validated against to cultural heritage data corpora.
With regard to our requirements, \citet{zaveri2015} highlight link quality as a primary concern for LOD, yet they address the hybrid nature of semi-structured data only superficially and provide limited coverage of cultural heritage-specific needs, such as domain-informed notions of trustworthiness.

\emph{\citet{black2020}.} They propose a comprehensive catalog comprising a versatile set of dimensions intended for broad applicability.
While this breadth increases enhances general usability, it reduces specificity for semi-structured aspects, as assumptions about data representation remain high-level.
For example, the catalog does not differentiate between temporal and non-temporal consistency, nor does it distinguish compliance from consistency.
This conceptual ambiguity undermines the applicability of the catalog to semi-structured data in domains such as cultural heritage.
In relation to our requirements, the catalog offers wide coverage, but provides limited support for the mixed nature of semi-structuredness, and only partially addresses link quality, primarily through the correctness of data file linkage.

Table~\ref{tab:relatedWorkSummary} summarizes how existing key publications discussed above address the three central aspects of descriptive information about cultural heritage objects introduced at the beginning of this section.
It provides a structured comparison that highlights the degree of alignment between existing definitions and the requirements of the cultural heritage domain.

\begin{table}[htbp]
  \centering
  \caption{
    Summary of how the quality definitions address the characteristics (i)~-~(iii) of object description data.
    The symbols below indicate the degree to which each aspect is addressed: (+) fully addressed, (–) not addressed, \((\pm)\) partly addressed, \((\sim)\) addressed only on a general level.
  }\label{tab:relatedWorkSummary}
  \begin{tabularx}{\textwidth}{lYYY}
    \textbf{Publication}              &
    \textbf{(i)\,Semi-structuredness} &
    \textbf{(ii)\,Link quality}       &
    \textbf{(iii)\,Domain-specific aspects}                          \\ \toprule
    \citet{wand1996}                  & \(\pm\)  & \(-\)   & \(-\)   \\
    \citet{scannapieco2004}           & \(\sim\) & \(-\)   & \(-\)   \\
    ISO~25012~\citep{iso:25012}       & \(-\)    & \(-\)   & \(\pm\) \\
    \citet{batini2009}                & \(+\)    & \(-\)   & \(-\)   \\
    \citet{zaveri2015}                & \(\pm\)  & \(+\)   & \(\pm\) \\
    \citet{firmani2020}               & \(-\)    & \(-\)   & \(\pm\) \\
    \citet{black2020}                 & \(\sim\) & \(\pm\) & \(\pm\)
  \end{tabularx}
  \Description{Summary of how the quality definitions address the characteristics (i)~-~(iii) of object description data.
    The symbols indicate the degree to which each aspect is addressed: (+) fully addressed, (–) not addressed, \((\pm)\) partly addressed, \((\sim)\) addressed only at a general level.}
\end{table}

\emph{FAIR and CARE principles.}
In the context of best practices for data management, two widely recognized sets of principles have gained prominence in discussions of DQ: FAIR and CARE.
The FAIR guiding principles by \citet{wilkinson2016} aim to improve the transparency, reproducibility, and reusability of research data by ensuring that it is findable, accessible, interoperable, and reusable.
The CARE principles by \citet{carroll2020} are intended to complement FAIR by emphasizing collective benefit, authority of control, responsibility, and ethics, particularly in relation to data about or affecting indigenous peoples.
While the FAIR principles focus primarily on research data and the CARE principles on data for governance, both principles are highly cited in discussion of DQ within the cultural heritage domain.
Importantly, neither set of principles provides an explicit definition of DQ; rather, they present guidelines for improving it.

\begin{answerbox}[label={ans:rq1}]{Answer to~\ref{rq:related-work} (\emph{What existing quality definitions are applicable to descriptive information about cultural heritage objects?})}
  None of the existing quality definitions can be adopted in their entirety.
  Many are designed for structured data or for domains other than descriptive information about cultural objects.
  They all do not adequately capture the combined properties of constrained and unconstrained elements as they occur in semi-structured data.
  Aspects such as link quality, provenance, trustworthiness, traceability of changes, and the explicit representation of uncertainty are addressed only partially, if at all.
  While certain quality aspects, such as those discussed by \citet{zaveri2015} for LOD, represent progress toward addressing some of these aspects, they remain insufficient for fully capturing all the aspects of object description data.
  Therefore, existing definitions cannot be directly applied, rather, they provide aspects that can be adapted and used as a basis for a DQ definition tailored specifically to descriptive information about cultural objects.
\end{answerbox}

%% file: 4_0_data_quality_dimensions.tex
\section{Defining Data Quality for Descriptive Information about Cultural Heritage Objects}\label{sec:data_quality_dimensions}
The definitions of data quality (DQ) reviewed in Section~\ref{sec:related_work} exhibit substantial variation and not fully meet the specific requirements of the cultural heritage domain, because the definitions are not tailored to its data.
Accordingly, this section aims to define DQ specifically for descriptive information about cultural heritage objects, thereby addressing~\ref{rq:definition} (\emph{How should the quality of descriptive information about cultural heritage objects be defined?}).
Following the approach commonly found in the literature, we define DQ through a set of quality dimensions, with each dimension representing a distinct aspect of DQ.
Given the characteristics (i)~-~(iii) presented in Section~\ref{sec:related_work}, the dimensions we propose aim to capture the essential aspects of the cultural heritage domain, including linkage and provenance.

When defining DQ dimensions, we focus both on the intrinsic aspects of data and its relationship with other contexts.
Accordingly, we distinguish between two categories:
\emph{intrinsic} dimensions and \emph{contextual} dimensions.
This distinction ensures that the set of DQ dimensions encompasses not only the essential properties of the data but also relevant contextual aspects.
The dimensions are systematically organized according to their relationship with content, and individual dimensions may include subdimensions, which are treated as distinct quality dimensions while remaining subordinate or auxiliary to their parent dimensions.
Each dimension is presented in a uniform structure:
a precise definition, illustrative examples drawn from real-world cultural heritage datasets, and a discussion of the conceptual origins of the definition.
Two conceptual clarifications guide this approach.
First, we employ the notion of a \emph{domain of interest} to emphasize that different subdomains, such as architecture or painting, may entail specific requirements and perspectives, even within the broader field of cultural heritage.
Secondly, we consider the \emph{context of use}, recognizing that the same data may serve diverse purposes, such as academic research or public dissemination, and that these contexts influence the perception and prioritization of quality.
Our definitions are based on the terminology in Section~\ref{sec:basic_terms}.

By applying a consistent structure of definition, illustrative examples, and explanation of conceptual origins, the resulting set of quality dimensions is presented systematically, providing a rigorous foundation for defining the quality of descriptive data about cultural heritage objects.

\subsection{Intrinsic Dimensions}
\input{4_1_intrinsic.tex}

\subsection{Contextual Dimensions}
\input{4_2_contextual.tex}

%% file: 4_1_intrinsic.tex
We begin with the intrinsic dimensions, which concern properties inherent to the data itself.
This group comprises \emph{Accuracy}, \emph{Consistency}, \emph{Conciseness}, and \emph{Completeness}.
\\
\begin{dimensiondef}[Accuracy]
  \emph{Accuracy} captures the degree to which a dataset correctly and precisely represents information, in accordance with defined standards.
  \emph{Accuracy} encompasses several subdimensions: \emph{Syntactic Accuracy}, \emph{Semantic Accuracy}, \emph{External Accuracy}, \emph{Compliance}, and \emph{Precision}.
  \emph{Syntactic Accuracy} refers to the degree to which the dataset conforms to the syntax of the employed data description language and, where applicable, the associated data model.
  \emph{Semantic Accuracy} refers to the degree to which the dataset correctly represents information of the domain of interest.
  \emph{External Accuracy} refers to the degree to which all interlinks in a dataset are valid and refer to the intended data elements in external datasets.
  \emph{Compliance} refers to the degree to which the dataset conforms to rules and guidelines established for the employed data description language and, where applicable, the associated data model.
  \emph{Precision} of a dataset refers to the degree of exactness of recorded data values, evaluated with respect to the requirements of a particular context of use.
\end{dimensiondef}

\begin{examples}
  \item[Syntactic A.]
  In the cultural heritage context, LIDO data should conform to the XML data description language and to the LIDO data model to be syntactically correct.
  \item[Semantic A.]
        (1) The Mona Lisa is a painting by Leonardo da Vinci.
  Attributing the painting to the artist Pablo Picasso would be semantically inaccurate.
  (2) If a building can be attributed with a date, it depends on the description in the data model what kind of date it is.
  For instance, recording the foundation date of a church when the model specifies the completion date would be semantically inaccurate.
  \item[External A.]
        (1) A place of birth may be indicated by a name, which should ideally be accompanied by a link to a location register.
  For external accuracy, this link must be valid and resolvable.
  (2) A link to an artwork type is compact if the type is as specific as possible.
  Typical registers such as the controlled vocabulary Art \& Architecture Thesaurus (AAT)~\citep{getty-vocabularies} capture multidimensional relationships and provide links to general and more specific entries.
  \item[Compliance]
        (1) The internationally recognized ISO~8601~standard~\citep{iso:8601} prescribes the date format YYYY-MM-DD.
  (2) The guidelines for recording cultural heritage data may require photographs’ dimensions to be given in \texttt{cm}.
  \item[Precision]
  The required level of precision depends on the object and context.
  The height of the Burj Khalifa, for instance, can typically be reported in meters, which is sufficient to identify it as the tallest building.
  In contrast, the dimensions of a print sheet are commonly recorded in millimeters, often supplemented by precise plate measurements for greater accuracy.
\end{examples}

\begin{origin}
  \item[Syntactic A.]
  Our definition is based on the concept of \citet{batini2009} and the ooperationalization of \citet{zaveri2015} for Linked Open Data (LOD).
  While the formulation by \citet{batini2009} is general and not fully suited to semi-structured data, the definition by \citet{zaveri2015} is LOD-specific.
  Accordingly, our approach builds on these approaches by incorporating alignment with the underlying data model and description language, addressing quality issues that are characteristic of object description data, where schema flexibility often fosters syntactic inaccuracy.
  \item[Semantic A.]
  Our definition is grounded in the widely recognized notion of \enquote{matching real-world facts}, as noted by \citet{wand1996}, ISO~25012~\citep{iso:25012}, and Zaveri~\citep{zaveri2015}.
  While these works identify this principle, they do not elaborate on \enquote{real-world facts}, particularly with respect to the source and contextual framing.
  Our definition of \emph{Semantic Accuracy} explicitly incorporates the origin of the information as well as its context of use and domain of interest, addressing semantic accuracy in semi-structured data.
  \item[External A.]
  As prior works address aspects of link-related quality, none defines a dedicated dimension corresponding to  \emph{External Accuracy}.
  Zaveri et al.~\citep{zaveri2015} emphasize link completeness, whereas Black et al.~\citep{black2020} combine link correctness, with link presence and accuracy in linkages.
  Our definition is based on empirically observed, community-driven quality issues, emphasizing the correctness and validity of links to external resources.
  We consider the mere existence or absence of such links, under the dimension of \emph{Completeness}.
  \item[Compliance]
  While ISO~25012~\citep{iso:25012} and \citet{black2020} define `compliance' applicable to semi-structured data, both definitions rely on generalized language and semantics.
  We have adapted these definitions to explicitly emphasize inherent aspects and incorporate semi-structured characteristics such as data description language.
  \item[Precision]
  \citet{wand1996} first addressed precision, albeit only by negating the notion of \enquote{inaccuracy}.
  Although ISO~25012~\citep{iso:25012} and \citet{black2020} align more closely with our understanding, ISO~25012~\citep{iso:25012} remains overly general, whereas \citet{black2020} provide greater specificity.
  Consequently, we adpopt semantic framing by \citet{black2020} and adapt the terminology to capture semi-structured characteristics.
\end{origin}

\begin{dimensiondef}[Completeness]
  \emph{Completeness} refers to the extent to which all information required for a particular context of use is present.
  \emph{Completeness} includes \emph{Internal} and \emph{External Completeness}.
  \emph{Internal Completeness} refers to the degree to which a dataset contains all data elements of the domain of interest required for a particular context of use, and each of those data elements includes the properties and links required for the intended context of use.
  \emph{External Completeness} refers to the degree to which a dataset contains all interlinks to external datasets that are required for a particular context of use.
\end{dimensiondef}

\begin{examples}
  \item[Internal C.]
        (1) In a register of paintings, all paintings created by a particular artist must be included to create a chronological overview of the artist’s work.
  (2) To determine the century in which a person lived, the year of birth may be sufficient.
  However, assigning a zodiac sign or software-based verification to disambiguating persons with the identical names typically requires the full date of birth.
  \item[External C.]
  To avoid ambiguity, each place of birth should be indicated by a valid link to a specific location register, such as GeoNames~\citep{geonames}.
\end{examples}

\begin{origin}
  \item[Internal C.]
  This dimension appears frequently in the literature.
  Accordingly, we build on the definitions provided by \citet{wand1996}, \citet{scannapieco2004}, \citet{black2020}, ISO~25012~\citep{iso:25012}, and \citet{batini2009}.
  Because these existing definitions do not distinguish between completeness at the dataset level and at the level of individual data elements, we introduce a more refines formulation
  \citet{zaveri2015} explicitly articulates this distinction, we adopt and adapt their approach to capture the internal structure of semi-structured datasets more accurately.
  \item[External C.]
  The ISO~25012~standard~\citep{iso:25012} delineates the concept of `completeness' through the utilization of \enquote{related entity instances}, a notion that can be interpreted as encompassing interlinking.
  \citet{zaveri2015} developed the term \enquote{linking completeness} for LOD as the first explicit articulation of this concept.
  We follow their interpretation, but replace the notion of \enquote{instances} with terminology suited to semi-structured data.
  Given the central role of interlinking in semi-structured data, we distinguish between \emph{External Completeness} and \emph{Internal Completeness} to capture the full range of completeness-related aspects.
\end{origin}

\begin{dimensiondef}[Conciseness]
  \emph{Conciseness} reflects the degree to which a dataset presents essential information efficiently.
  \emph{Conciseness} comprises \emph{Relevance} and \emph{Compactness}.
  \emph{Relevance} refers to the degree to which a dataset contains only those data elements, links and properties required for a particular context of use.
  \emph{Compactness} refers to the degree to which a dataset contains no redundant elements and all elements are compact.
  A data element is compact if it does not contain redundant links or properties.
  A data value is compact if it is as condensed as possible.
\end{dimensiondef}

\begin{examples}
  \item[Relevance]
  In the metadata for a painting, specifying the artist of the painting is relevant.
  In contrast, including extraneous personal information about the artist, such as details of their clothing, is irrelevant for the purpose of describing the artwork.
  \item[Compactness]
        (1) A dataset of paintings from an exhibition is compact if it does not contain multiple elements describing the same painting.
  (2) A data element that specifies the size of a photograph is compact if it includes only the relevant measurement (e.g., the size of photograph itself) without redundant information (e.g., the frame size, unless contextually required).
  (3) A data value for the width of a photograph is compact if it is expressed in a concise numeric form, without unnecessary leading zeros.
\end{examples}

\begin{origin}
  \item[Relevance]
  Our definition of \emph{Relevance} aligns with the definitions presented by \citet{black2020} and \citet{zaveri2015}.
  However, the definition by \citet{black2020} is overly broad for our purposes, and \citet{zaveri2015} define `completeness' as the presence of relevant information in the context of use.
  In contrast, our definition emphasizes both the degree to which present information is relevant and the absence of irrelevant information.
  Our definition also differentiate between elements, properties, and links, rather than using the generic term \enquote{information}.
  \item[Compactness]
  The ISO~25012~standard~\citep{iso:25012} addresses compactness indirectly under efficiency, while \citet{zaveri2015} consider it within the dimension of `representational conciseness', emphasizing how data is represented rather than the intrinsic compactness of the data content.
  \citet{black2020} approach the aspect of compactness via `redundancy' and `uniqueness', but their focus is limited to duplicate detection and does not fully capture the compactness of information.
  In line with \citet{zaveri2015}, we distinguish compactness from relevance, explicitly defining \emph{Compactness} as an independent dimension that addresses the efficient, non-redundant representation of data elements, properties, and values.
\end{origin}

\begin{dimensiondef}[Consistency]
  \emph{Consistency} refers to the uniformity of information within a dataset without internal conflicts.
  \emph{Consistency} comprises \emph{Logical Consistency} and \emph{Coherence}.
  \emph{Logical Consistency} refers to the degree to which a dataset is free from logical contradictions within a given domain of interest and context of use.
  \emph{Coherence} refers to the degree to which the dataset contains coherent data elements.
  Two data elements are coherent if their respective properties and links are coherent in pairs.
  Two properties are coherent if their data values are the same when representing the same information.
  Two links are coherent if they refer to the same element when representing the same information.
  Two data values are coherent if similar information is expressed in a consistent format.
\end{dimensiondef}

\begin{examples}
  \item[Logical Consistency]
  The date of a person’s death must not precede their date of birth; any violation constitutes a logical contradiction.
  \item[Coherence]
        (1) In a dataset of photograph metadata, all elements are coherent if measurements for width and height are specified consistently across records.
  (2) Width and height values should use the same units, such as \texttt{cm}, for all elements.
  (3) Location name \texttt{London} must be unambiguous for cities;
  This can be achieved, e.g., by a suffix such as \texttt{London~(United Kingdom)}.
  (4) Each city should be represented by a single data element, ensuring that all links to that city are unambiguous.
  (5) Measurement values are coherent if their formats are identical.
  For example, integers should be consistently represented without leading zeros.
\end{examples}

\begin{origin}
  \item[Logical Consistency]
  ISO~25012~\citep{iso:25012} aligns with our understanding of \emph{Logical Consistency}, but they combine the aspects of \emph{Coherence} and \emph{Logical Consistency} into a single, overly broad dimension.
  \citet{zaveri2015} provide a more formal, rule-based definition, yet it only partially addresses logical contradictions.
  \citet{black2020} also focus on rule-based aspects, which we associate more appropriately with our definition of \emph{Compliance}.
  Accordingly, we explicitly define \emph{Logical Consistency} as a distinct dimension, emphasizing logical contradictions as an inherent quality issue in semi-structured data.
  \item[Coherence]
  While \citet{wand1996} address `Coherence' in terms of information system states, our understanding aligns more closely with \citet{scannapieco2004} and \citet{batini2009}, though we adopt a more semantic- and rule-based perspective.
  To account for the characteristics of semi-structured data, our definition explicitly targets \emph{Coherence} across datasets, elements, properties, and values.
\end{origin}

%% file: 4_2_contextual.tex
Next, we consider the contextual dimensions, which capture aspects of data quality that depend on the broader context in which the data is used or managed.
\emph{Accessibility} pertains to availability of data within a processing environment, \emph{Currency} and \emph{Traceability} refer to the management of different data versions, \emph{Plausibility} addresses the credibility of data, and \emph{Understandability} refers to the ease with which users can process data.
\\
\begin{dimensiondef}[Accessibility]
  \emph{Accessibility} refers to additional information that govern access to data.
  \emph{Accessibility} comprises \emph{Availability}, \emph{Confidentiality}, and \emph{Integrity} of data.
  \emph{Availability} is the degree to which data can be accessed when required for a given context of use.
  \emph{Confidentiality} refers to the restriction of access to authorized users only, ensuring that sensitive data is protected according to the requirements of a given context of use.
  \emph{Integrity} refers to the degree to which data remains free from unintentional or unauthorized alterations, preserving its intended state over time.
\end{dimensiondef}

\begin{examples}
  \item[Availability]
  Accessing a resource through a website enhances its availability.
  Specifying a unique and persistent identifier, such as a Digital Object Identifier (DOI), further improves availability by ensuring that an organization is responsible for maintaining the resource's integrity and long-term accessibility.
  \item[Confidentiality]
        (1) Certain information, such as the owner’s name of an object form, may be crucial for provenance tracking in a museum but must be hidden during data export to protect privacy.
  (2) Role-based access control is often applied to data collections.
  For example, new staff may be permitted to edit newly entered data, whereas editors have write access to all the data within a collection.
  \item[Integrity]
  Techniques such as digital signatures and cryptographic hashes can be employed to verify the authenticity of data changes, thereby distinguishing authorized changes from unauthorized or accidental alterations.
\end{examples}

\begin{origin}
  \item[Availability]
  We have refined prior work to develop our definition of \emph{Availability}.
  Unlike the ISO 25012 standard~\citep{iso:25012}, we explicitly separate authorization from confidentiality.
  Unlike \citet{zaveri2015}, we prioritize actual access rather than mere presence, as presence is addressed under our dimensions of \emph{Completeness}.
  In contrast to \citet{black2020}, who use the terms \enquote{consulted} and \enquote{retrieved}, we adopt \enquote{access} as a more appropriate term and incorporate the context of use.
  \item[Confidentiality]
  Our definition of \emph{Confidentiality} builds on prior work.
  From ISO~25012~\citep{iso:25012}, we adopt the notion that access is context-dependent and restricted to authorized users.
  From the definition by \citet{zaveri2015}, we incorporate the crucial distinction between access data and the granting of rights, which underpins our approach.
  Compared to \citet{black2020}, who emphasize restricting access to authorized users, we highlight the interplay between confidentiality and general accessibility.
  \item[Integrity]
  Our definition of \emph{Integrity} differs from those in ISO~25012~\citep{iso:25012}, \citet{zaveri2015}, and \citet{firmani2020}, which focus primarily on data protection and security, and from \citet{black2020}, which focus on \enquote{data value loss} and \enquote{corruption}.
  Our definition takes a broader perspective, encompassing all forms of unintended data alteration, and thereby extending the concept's applicability to semi-structured datasets.
\end{origin}

\begin{dimensiondef}[Currency]
  \emph{Currency} refers to the extent to which a dataset reflects up-to-date information from the domain of interest.
  \emph{Currency} comprises \emph{Data Currency}, \emph{Data Update Currency}, and \emph{Time Concurrency}.
  \emph{Data Currency} refers to the degree to which a dataset contains up-to-date information relevant to the domain of interest.
  \emph{Data Update Currency} refers to the degree to which a dataset is updated promptly after information in a domain of interest has changed.
  \emph{Time Concurrency} assesses whether all data elements, including their values and links, refer consistently to the same version of information in the domain.
\end{dimensiondef}

\begin{examples}
  \item[Data C.]
  Paintings can change ownership at auctions.
  To accurately determine the current location of an artist’s paintings, the data must always be up-to-date.
  \item[Data Update C.]
        (1) Museum inventories are conducted periodically and may reveal discrepancies with existing data.
  The closer an inventory is performed relative to changes in holdings, the more current the data.
  (2) Natural dynamics, such as the tilt of the Leaning Tower of Pisa, are monitored regularly.
  The more recent the measurements, the more accurately the data reflects the current state.
  \item[Time Concurrency]
  In a museum, paintings may be relocated as part of an exhibition.
  If some paintings' location data are already updated to the current exhibition, while others remain outdated, the data lacks time concurrency, resulting in inconsistent representation of the exhibition.
\end{examples}

\begin{origin}
  \item[Data C.]
  We based our definition of \emph{Data Currency} closely on prior work.
  Our definition aligns with the definition of `currentness' by ISO~25012~\citep{iso:25012} (referred to there as \enquote{right age}), with the definition of `timeliness' by \citet{zaveri2015}, and with the definition of `currency' by \citet{black2020}.
  We define \emph{Data Currency} to emphasize the importance of currency in the data itself, and to explicitly integrate the notion of domain of interest, since \enquote{up-to-date} is inherently domain-specific.
  \item[Data Update C.]
  Our definition of \emph{Data Update Currency} focuses on the temporal lag between changes in the real world and their reflection in the dataset.
  While \citet{wand1996} and \citet{scannapieco2004} address `timeliness', they frame it in terms of the information system rather than the domain.
  Our approach isolates this dimension from the system context.
  \item[Time Concurrency]
  In cultural heritage, datasets typically describe multiple objects, which may evolve asynchronously in different records.
  This dimension remains unaddressed in the literature we selected, but represents a recurring challenge in cultural heritage data management \citep{ferradji2022}.
\end{origin}

\begin{dimensiondef}[Plausibility]
  \emph{Plausibility} refers to the degree to which information can be considered credible.
  \emph{Plausibility} comprises \emph{Provenance} and \emph{Trustworthiness} of data.
  \emph{Provenance} refers to the degree to which a dataset, including its elements and values, are equipped with verifiable information about its origin.
  \emph{Trustworthiness} refers to the degree to which a dataset, including its elements and values, are believed by a relevant community within a specific context of use.
\end{dimensiondef}

\begin{examples}
  \item[Provenance]
  A death certificate cited as the source for a person’s date of death provides verifiable evidence, enhancing the plausibility of the data.
  \item[Trustworthiness]
  Cultural heritage information published by a reputable museum about an object in its possession is generally more trustworthy than information in tourist promotional material, as the museum is recognized for its expertise and responsible handling of information.
  Similarly,  the reputation, education, and experience of the data contributor influence trustworthiness.
\end{examples}

\begin{origin}
  \item[Provenance]
  We define \emph{Provenance} by reviewing prior concepts: Firmani et al.~\citep{firmani2020} emphasize transparency through verifiable origins, process and metadata.
  \citet{black2020} define `traceability' by focusing on technical lineage and define `reputation' by focusing on trustworthiness.
  Our definition refers to the origin of information as an inherent aspect, a critical yet underrepresented factor in existing quality definitions that is highly important in research and cultural heritage, where accurate provenance underpins data reuse and scholarly reliability.
  \item[Trustworthiness]
  We developed the concept of \emph{Trustworthiness} by building on the following definitions:
  ISO~25012~\citep{iso:25012} addresses credibility in general, we adopt it to object description data; the definition by \citet{zaveri2015} omits the perspective of a group of people; the definition of objectivity by \citet{black2020} neglects subjectivity, whereas we address the aspects of plausibility presented in \citep{black2020} in our dimensions of \emph{Accuracy}.
  Our definition focuses on the reliability of information, which is a critical yet underrepresented factor in existing quality definitions.
  However, it is important in research and cultural heritage, where confidence in information is essential for scholarly use.
\end{origin}

\begin{dimensiondef}[Traceability]
  \emph{Traceability} refers to the availability of information regarding changes of data over time.
  \emph{Traceability} comprises \emph{Temporal} and \emph{Causal Traceability}.
  \emph{Temporal Traceability} is the degree to which a dataset has an audit trail of access and a history of changes within a particular context of use.
  \emph{Causal Traceability} is the degree to which causal dependencies among dataset updates are documented and accessible for a given context of use.
\end{dimensiondef}

\begin{examples}
  \item[Temporal T.]
  Historic buildings are typically dynamic in nature.
  All changes, additions, and demolitions to a building should be recorded in a detailed history of change to maintain temporal traceability.
  \item[Causal T.]
  Changes to a historic building may trigger updates to related data, such as the group of architects responsible for modifications of the building.
  In this context, sculptures may have been relocated due to change.
\end{examples}

\begin{origin}
  \item[Temporal T.]
  Our definition of \emph{Temporal Traceability} is based on prior work: it draws on ISO~25012~\citep{iso:25012}, but we have adapted the wording for clarity.
  The definition of traceability by \citet{black2020} is similar, but focuses narrowly on technical data lineage.
  Our approach encompasses both technical and contextual concepts, which are essential in the cultural heritage.
  \item[Causal T.]
  We introduced \emph{Causal Traceability} to address a gap in existing definitions, such as those outlined in ISO~25012~\citep{iso:25012} and \citet{black2020}, which focus on temporal aspects of traceability but neglect causal relationships.
  \citet{firmani2020} include the interpretation of information in their definition of transparency, but their focus differs from ours: they emphasize explaining a specific extraction result, whereas we emphasize reconstructing the causal chain of dataset updates.
  In cultural heritage, understanding how and why information was created is essential to accurately interpreting, reusing, and conducting research on it \citep{hannah2014}.
  Our definition explicitly captures these causal aspects, providing deeper insight into traceability and the contextual data factors.
\end{origin}

\begin{dimensiondef}[Understandability]
  \emph{Understandability} refers to the ease with which data can be comprehended and processed by users, including both humans and tools.
  \emph{Understandability} comprises \emph{Appropriateness} and \emph{Versatility}.
  \emph{Appropriateness} refers to the ease with which data can be understood without ambiguity within a given context of use.
  \emph{Versatility} refers to the extent to which data is available in different forms across different cultural perspectives, technology platforms, and user sensory abilities.
\end{dimensiondef}

\begin{examples}
  \item[Appropriateness]
  The city name ``London'' as a birthplace may be ambiguous and can lead to misunderstandings, as there are multiple cities with that name in different countries.
  Providing geolocations or references to a location register reduces ambiguity and increase understanding.
  \item[Versatility]
        (1) Measurements in the metric system are easily understood in Europe but may be less intuitive for US Americans.
  (2) The date of New Year’s celebration varies between the Gregorian and lunar calendars, necessitating contextual clarification.
  (3) Historical dates prior to 1582 require distinguishing between the Julian and Gregorian calendars.
  Providing data in multiple formats or adhering to ISO~8601~\citep{iso:8601} can prevent misinterpretation and enhance versatility.
\end{examples}

\begin{origin}
  \item[Appropriateness]
  Our definition of \emph{Appropriateness} is based on related concepts.
  Both ISO~25012~\citep{iso:25012} and \citet{zaveri2015} emphasize ease of comprehension and contextual interpretation.
  While our definition largely aligns with these two, the definition of ISO~25012~\citep{iso:25012} focuses more on quantitative aspects.
  \item[Versatility]
  Our definition of \emph{Versatility} synthesizes aspects from related definitions.
  The definition of `accessibility' in the ISO~25012~standard~\citep{iso:25012} acknowledges varying user needs, but focuses on access rather than comprehension.
  The definition of `interpretability' by \citet{zaveri2015} addresses technical processing and only partially overlaps with our focus.
  The definition of `social diversity' by \citet{firmani2020} highlight multiple representations:
  We refine these perspectives into a cohesive definition emphasizing adaptability across diverse contexts of understanding, reflecting the practical needs of the cultural heritage domain.
\end{origin}

\begin{answerbox}[label={ans:rq2}]{Answer to~\ref{rq:definition} (\emph{How should the quality of descriptive information about cultural heritage objects be defined?})}
  Building on a clear conceptual foundation, we first defined the basic terms necessary to address the quality of descriptive information about cultural heritage objects.
  We then defined the quality of such data through a comprehensive set of quality dimensions that capture its characteristics:
  (i)~semi-structuredness (addressed through \emph{Syntactic}/\emph{Semantic}/\emph{External Accuracy}, \emph{Compliance}, \emph{Completeness}, \emph{Coherence}, and \emph{Precision}), (ii)~link quality (addressed via \emph{External Accuracy} and \emph{Completeness}, \emph{Availability}, and \emph{Integrity}), and (iii)~domain-specific aspects (captured through \emph{Provenance}, Trustworthiness, Temporal/\emph{Causal Traceability}, \emph{Currency}, and \emph{Understandability}).
  This dimension-based definition is illustrated with real-world examples tailored to the cultural heritage and grounded in a structured process that traces the conceptual origins of each dimension.
  The proposed framework explicitly considers both the domain of interest and the context of use, ensuring that the definition is sensitive to the specific requirements and purposes of object description data.
\end{answerbox}

%% file: 5_0_empirical_evaluation.tex
\section{Empirical Evaluation}\label{sec:empirical_evaluation}
Data quality (DQ) dimensions serve to classify DQ problems and guide data assurance techniques with in specific domains.
They also define the aims of quality assurance by indicating which aspects of data must be ensured.
In this section, we evaluate our dimension-based DQ definition using a catalog of empirically collected quality problems from the cultural heritage domain (cf.~\citet{kesper2023}).
The catalog was compiled for a community and dataset that align exactly with our target scope.
To answer~\ref{rq:evaluation} (\emph{Is the proposed data quality definition for descriptive information of cultural heritage objects appropriate?}) we subdivide the analysis into three subquestions, each targeting a distinct aspect of the proposed quality dimensions.
\ref{rq:evaluation:completeness} examines whether the framework of dimensions is sufficient to classify the full range of quality problems, thereby assessing its completeness.
\ref{rq:evaluation:necessity} investigates whether all defined dimensions are practically necessary for classification, helping to identify potential redundancies.
\ref{rq:evaluation:orthogonality} evaluates whether the dimensions are orthogonal (i.e., conceptually distinct from another) to ensure that the framework avoids overlap and redundancy.
Taken together, the following subquestions provide a systematic way to evaluating the relevance, sufficiency, and conceptual soundness of the quality dimensions defined:

\begin{enumerate}[label= \textbf{RQ3\alph*}, leftmargin=*, topsep=6pt]
  \item\label{rq:evaluation:completeness} Can all quality problems be classified according to the proposed quality dimensions?
  \item\label{rq:evaluation:necessity} Are all dimensions necessary for classifying the collected set of quality problems?
  \item\label{rq:evaluation:orthogonality} Are the proposed quality dimensions orthogonal to each other?
\end{enumerate}

\subsection{Set up}
\input{5_1_setup.tex}

\subsection{Assignment of quality problems to quality dimensions}
\input{5_2_assignment_quality_problems.tex}

\subsection{Orthogonality of quality dimensions}
\input{5_3_orthogonality_quality_dimensions.tex}

\subsection{Threats to Validity}
\input{5_4_threats_to_validity.tex}

%% file: 5_1_setup.tex
For our evaluation, we use a catalog~\citep{kesper2023} of 51 distinct DQ problems from the cultural heritage domain.
Each problem in the catalog is documented in a structured way by noting various aspects, such as its impact on DQ and potential causes.
The identification of problems was based on an empirical study, with each problem checked for its occurrence in at least two representative datasets and one corresponding data model.
The study was conducted through a community workshop involving 19 participants from galleries, libraries, archives, and museums (GLAM institutions), complemented by six expert interviews.
The workshop employed an explorative World Café format \citep{fouche2011}, while the expert interviews followed established qualitative research guidelines \citep{bogner2014}.
All empirical problems were validated by testing incidents in two datasets provided by the Scientific Collections of the Georg-August-University Göttingen~\citep{goettingen_collections} and the German Documentation Center for Art History~\citep{ddk_uni_marburg}.
These datasets, described in detail in the referenced catalog, encompass a broad spectrum of descriptive information about cultural heritage objects.
Quality problems were verified through a combination of manual inspection and pattern-based sampling using XQuery and XSLT.
In the original catalog, quality problems were classified using the term \emph{quality features}.
In this work, we adopt the term \emph{subdimensions} while preserving the original conceptual meaning.

%% file: 5_2_assignment_quality_problems.tex
To address RQ3a and RQ3b, the catalog~\citep{kesper2023} classifies each quality problem according to the affected quality dimensions, distinguishing between primary and other affected dimensions.
For each assignment, a justification is provided based on the definitions of the respective quality dimensions.
In cases where a quality dimension was only partially affected, the relevant conditions were explicitly specified.
When a quality problem has multiple causes or variants, the extent to which a quality dimension is affected depends on the specific cause or variant.
For example, problem DATA06.2 (Imprecision) pertains to statements in the data that do not meet context-specific requirements for \emph{Precision}.
In addition to \emph{Precision}, \emph{Appropriateness} may also be affected when the range of information, such as \enquote{around 1900}, is ambiguous.
The complete categorization of quality problems by dimension is provided in the appendix (cf. Table~\ref{tab:dataQualityProblems}).
Table~\ref{tab:distDQDimensions} summarizes the frequency with which each quality dimension is identified as the primary dimension affected in the catalog.

\begin{table}[ht]
  \centering
  \caption{List of affected data quality dimensions}\label{tab:distDQDimensions}
  \footnotesize
  \begin{minipage}[t]{0.48\textwidth}
    \centering
    \textbf{Intrinsic Dimensions}\\[0.5ex]
    \begin{tabular}{lrr}
      \toprule
      \textbf{Dimension}    & \textbf{Count} & \textbf{\%} \\
      \midrule
      Syntactic Accuracy    & 12             & 6.49\%      \\
      Semantic Accuracy     & 18             & 9.73\%      \\
      External Accuracy     & 8              & 4.32\%      \\
      Compliance            & 25             & 13.51\%     \\
      Precision             & 8              & 4.32\%      \\
      Internal Completeness & 21             & 11.35\%     \\
      External Completeness & 15             & 8.11\%      \\
      Relevance             & 0              & 0.00\%      \\
      Compactness           & 4              & 2.16\%      \\
      Logical Consistency   & 6              & 3.24\%      \\
      Coherence             & 12             & 6.49\%      \\
      \bottomrule
    \end{tabular}
  \end{minipage}%
  \hfill
  \begin{minipage}[t]{0.48\textwidth}
    \centering
    \textbf{Contextual Dimensions}\\[0.5ex]
    \begin{tabular}{lrr}
      \toprule
      \textbf{Dimension}    & \textbf{Count} & \textbf{\%} \\
      \midrule
      Availability          & 3              & 1.62\%      \\
      Confidentiality       & 0              & 0.00\%      \\
      Integrity             & 0              & 0.00\%      \\
      Data Currency         & 3              & 1.62\%      \\
      Data Update Currency  & 1              & 0.54\%      \\
      Time Concurrency      & 3              & 1.62\%      \\
      Provenance            & 4              & 2.16\%      \\
      Trustworthiness       & 8              & 4.32\%      \\
      Temporal Traceability & 3              & 1.62\%      \\
      Causal Traceability   & 2              & 1.08\%      \\
      Appropriateness       & 24             & 12.97\%     \\
      Versatility           & 5              & 2.70\%      \\
      \bottomrule
    \end{tabular}
  \end{minipage}
\end{table}

\emph{Findings}.
The full categorization of quality problems by dimension (cf. Table~\ref{tab:dataQualityProblems}) demonstrates that each problem is assigned to at least one quality dimension.
This categorization provides an explicit rationale for each assignment, which is explained in detail in the catalog~\citep{kesper2023} itself.
As summarized in Table~\ref{tab:distDQDimensions}, a single problem is associated, on average, with 3.5 quality dimensions (including both primary and other assignments).
The most frequently assigned quality dimensions are \emph{Compliance} \(13,51\%\) and \emph{Appropriateness} \(12,97\%\).
Other quality dimensions that occur quite frequently include \emph{Internal} and \emph{External Completeness}, \emph{Syntactic} and \emph{Semantic Accuracy}, \emph{Trustworthiness}, and \emph{Coherence}.
In contrast, three dimensions were not assigned to any of the problems in the catalog: \emph{Relevance}, \emph{Confidentiality}, and \emph{Integrity}.

\emph{Interpretation}.
RQ3a can be answered as follows: all problems in the catalog can be meaningful assigned to the proposed quality dimensions, and each assignment is supported by a defensible rationale.
The frequent affection of \emph{Compliance} and \emph{Appropriateness} can be explained by their particular relevance in the cultural humanities.
Domain experts regularly rely on controlled vocabularies to ensure conformity of information, and they often must express forms of uncertainty in a transparent and intelligible way.
A similar explanation applies to the frequent affection for completeness and accuracy:
experts in cultural humanities place strong emphasis on collecting as much information as possible and documenting it with a high degree of correctness and precision.
As cultural heritage data is often intended to be read by humans, experts also attach considerable importance to the appropriateness of the data representation.

In response to RQ3b, several quality dimensions were not required for classifying any of the quality problems in the catalog.
The absence of assignments to the DQ dimension \emph{Relevance} likely reflects the domain experts' aim to achieve maximal completeness.
Since the curated repositories are intended for broad and heterogeneous use cases, all captured information is implicitly assumed to be relevant.
\emph{Data Currency}, \emph{Data Update Currency}, and \emph{Time Concurrency} were rarely assigned, presumably because these dimensions are more characteristic of other types of data, such as measurement data, than of object description data.
The lack of problems related to \emph{Confidentiality} can be attributed to the strong open access and open science orientation in the cultural heritage domain.
Because data is generally meant for unrestricted public dissemination, confidentiality concerns are addressed prior to publication and therefore play little role in experts’ quality assessments.
\emph{Integrity} is also of minor importance for object description data.
As the curating organizations are primarily research-oriented, they generally do not consider adversial threats, and potential corruption would not have significant economic or political consequences.
Finally, the \emph{Temporal} and \emph{Causal Traceability} dimensions are of little importance in this context.
Domain experts rarely identify quality problems related to these types of traceability, and tend not to distinguish between them in practice, which likely explains their absence in the catalog.
% \todo[inline, color=cyan]{We could add, that these assessment matches with the view of multiple domain experts, with whom we discussed about that}
% \todo[inline, color=green]{Hmm. Können wir das belegen?}

%% file: 5_3_orthogonality_quality_dimensions.tex
To address RQ3c, we analyzed potential dependencies between the quality dimensions based on the classification of quality problems.
For this purpose, we applied the chi-square test for independence as described by~\citet{mittag2014}.
The chi-squared test is a statistical hypothesis test used to assess whether two categorical variables are independent.
In this analysis, each quality dimension is treated as a binary categorical variable indicating whether it is assigned to a given problem.

This analysis uses Equation~(\ref{eq:chi-square}) to compute the chi-squared statistic ($\chi^2$) for each pair of quality dimensions $(a,b)$ based on the observed frequencies $O_{ij}$ and the expected frequencies $E_{ij}$.
The expected frequencies are calculated under the null hypothesis that the two dimensions are independent, using Equation~(\ref{eq:chi-square-expected}) and Table~\ref{tab:distDQDimensions}.
Since the catalog contains 51 problems, chi-square values theoretically range from 0 to 51.
The null hypothesis states that each pair of dimensions is orthogonal, meaning that no dependency exists between them.
According to the chi-squared distribution with a significance level of 1\%, values exceeding 6.63 indicate statistically significant deviation from independence.

\begin{minipage}{0.48\linewidth}
\begin{equation}
\label{eq:chi-square}
\chi^2_{a,b} = \sum_{i\in\{a,\not a\}}
\sum_{j\in\{b,\not b\}}
\frac{(O_{ij}-E_{ij})^2}{E_{ij}}
\end{equation}
\end{minipage}\hfill
\begin{minipage}{0.48\linewidth}
\begin{equation}
\label{eq:chi-square-expected}
E_{ij} = \frac{O_i O_j}{O_{total}}
\end{equation}
\end{minipage}

\emph{Findings}.
Table~\ref{tab:chi2DataQualityDimensions} in the appendix lists the resulting chi-squared values for all dimension pairs.
Quality dimensions that were not be assigned to any quality problem were excluded, as chi-square statistics cannot be computed for categories with zero marginal frequencies.
This applies to \emph{Relevance}, \emph{Confidentiality}, and \emph{Integrity}.
Overall, all examined pairs exhibit values that indicate at least some degree of independence, none of the pairs approaches the maximum possible value.
We expected this outcome, as no pair of dimensions is consistently assigned together, or consistently not assigned, across the entire problem catalog.
Using a significance level of 1\%, the 10 dimension pairs with the highest values are as follows:
(\emph{Causal Traceability}, \emph{Temporal Traceability}) (33.31),
(\emph{Trustworthiness}, \emph{Provenance}) (23.33),
(\emph{Data Currency}, \emph{Data Update Currency}) (16.32),
(\emph{External Completeness}, \emph{External Accuracy}) (9.50),
(\emph{Time Concurrency}, \emph{Logical Consistency}) (9.26),
(\emph{Trustworthiness}, \emph{Compliance}) (9.12),
(\emph{External Completeness}, \emph{Internal Completeness}) (9.07),
(\emph{Appropriateness}, \emph{Compliance}) (8.6),
(\emph{Availability}, \emph{External Completeness}) (7.7), and
(\emph{Causal Traceability}, \emph{Time Concurrency}) (7.3).

\emph{Interpretation}.
The highest chi-squared values can largely be explained by the definitions of the corresponding dimensions and their low overall occurrence.
Several of the strongest associations involve dimension pairs with a very small marginal frequencies, such as:
(\emph{Causal Traceability}, \emph{Temporal Traceability}), (\emph{Data Currency}, \emph{Data Update Currency}), (\emph{Time Concurrency}, \emph{Logical Consistency}), and (\emph{Causal Traceability}, \emph{Time Concurrency}).
Similarly, the pairs (\emph{Availability}, \emph{External Completeness}) and (\emph{Trustworthiness}, \emph{Provenance}) include dimensions with very low frequencies.
Because low-frequency dimensions provide limited explanatory power, these results have little interpretative value and can be disregarded.
The chi-squared value of 9.12 for (\emph{Trustworthiness}, \emph{Compliance}) arises from the absence of problems associated with both dimensions.
This produces a strong deviation between observed and expected frequencies and suggests a potential negative dependency.
Although this contradicts the ideal of full orthogonality, we do not consider such negative dependencies as problematic: the dimensions describe conceptually distinct types of problems, and their seperation therefore supports the validaty of the quality definition.
The same reasoning applies to (\emph{Appropriateness}, \emph{Compliance}).
The pair (\emph{External Completeness} and \emph{External Accuracy}) has a chi-square value of 9.50.
This relationship is plausible, as problems affecting \emph{External Accuracy} often involve missing or incorrect links, thereby simultaneously reducing completeness.
Hence, partial dependency between the two dimensions is expected.
Finally, the pair \emph{External Completeness} and \emph{Internal Completeness} has a chi-square value of 9.07.
Problems affecting \emph{External Accuracy} often involve missing or incorrect links, which simultaneously reduce completeness.
Both are subdimensions of the same overarching quality dimension, and many cataloged problems are described at a coarse granularity.
Thus, this commonality is not only expected, but can also be logically explained by the problem catalog as domain experts did not always distinguish between internal and external completeness.
Although the dimensions are conceptually distinct, limitations in the granularity of the problem descriptions reduce the degree of empirical independence.
A more fine-grained problem catalog would likely yield lower chi-square values for this pair.

In response to RQ3c, we discovered that empirical independence holds for all but 10 of the 171 examined dimension pairs.
Of these ten pairs, six are based on low occurrence, and two reflect negative dependencies, which we do not consider problematic.
For the remaining two pairs, we have shown that the observed dependencies can be logically explained by the nature of the quality dimensions and the structure of the problem catalog.
Overall, no evidence was found that challenges the conceptual orthogonality of the proposed quality dimensions.

%% file: 5_4_threats_to_validity.tex
This section identifies several threats to validity in this study, distinguishing between internal, construct, and external validity.
\emph{Internal validity} may be affected by the manner in which the quality problems were collected.
\begin{itemize}
  \item Quality problems were elicited through interviews with domain experts.
        Although these experts represent diverse fields, institutions, and levels of experiences, the recorded problems remain inherently subjective.
  \item The resulting catalog of quality problems is empirical and cannot be considered exhaustive.
  \item The catalog of quality problems may also not be fully representative of object description data, as domain experts focussed on specific areas of interest.
        Issues such as heterogeneity in data representation, use of controlled vocabularies, and handling of uncertainty were frequently highlighted, whereas longstanding concerns, such as completeness and correctness of data, were mentioned less often.
\end{itemize}

\emph{Construct validity} could be threatened by the subjective nature of mapping quality problems to quality dimensions.
To mitigate this, a clear rationale for each problem assignment is provided in the accompanying catalog~\citep{kesper2023}.

\emph{External validity} may be limited by the choice of domain, databases, data models, and participating experts.
The evaluation was conducted within the cultural heritage domain, where data is primarily collected manually, introducing a degree of subjectivity in the data.
In addition, the domain expert's emphasis on open access and open science, combined with the high level of uncertainty inherent in the data may influence perceptions of DQ and the types of quality problems identified.
Future research should extend this evaluation to other domains, data types, and quality problem categories to assess the generalizability of the proposed quality dimensions.

\begin{answerbox}[label={ans:rq3}]{Answer to~\ref{rq:evaluation} (\emph{Is the proposed data quality definition for descriptive information of cultural heritage objects appropriate?})}
  In summary, the evaluation shows that our dimension-based definition of DQ is well-suited for descriptive information about cultural heritage objects.
  All empirically collected quality problems could be classified using the proposed dimensions, indicating that the set of dimensions is sufficiently comprehensive.

  Each defined dimension proved useful in practice, except for three that were not represented in the catalog:
  \emph{Relevance}, \emph{Confidentiality}, and \emph{Integrity}.
  The absence of \emph{Relevance}-related problems suggests that domain experts generally assume maximal completeness, treating all included information as inherently relevant.
  The lack of problems associated with \emph{Confidentiality} reflects the field’s openness orientation:
  data are intended for unrestricted public use, rendering this dimension less relevant for domain experts who are focused on the quality of shared datasets.
  Similarly, \emph{Integrity} is a minor concern for domain experts, who are primarily research-oriented, and do not encounter data corruption such as unintentional tampering.
  Furthermore, our analysis confirmed empirical independence for 161 out of 171 pairs of quality dimensions, confirming that the set of dimensions is largely orthogonal.
  Of the 10 non-independent pairs, six can be attributed to low sample size, two reflect negative dependencies that do not compromise orthogonality, and two represent conceptually explainable overlaps between related dimensions.
  As a result, our findings indicate that the proposed quality definition effectively captures the range of quality problems in the cultural heritage domain and provides a sound basis for systematic quality assessment.
\end{answerbox}

%% file: 6_conclusion.tex
\section{Conclusion}\label{sec:conclusion}
This work set out to examine how data quality (DQ) can be defined for descriptive information about cultural heritage objects.

The first step was to assess whether existing DQ definitions could be adopted in this domain.
Current approaches offer valuable dimensions, often tailored to structured data, information systems, or linked data.
However, our analysis revealed that these approaches only partially address the specific characteristics of descriptive information about cultural heritage objects.
Critical characteristics such as semi-structuredness, link quality, and domain-specific requirements remain insufficiently covered.
Consequently, we conclude that no existing definition can be directly adopted.
Rather, they offer valuable aspects that can serve as a basis for developing a domain-specific DQ definition.
To address this gap, we developed a comprehensive, dimension-based definition of DQ tailored to descriptive information about cultural heritage objects.
This definition accounts for the semi-structured nature of the data, the importance of linking across collections, and domain-specific requirements such as provenance, trustworthiness, and traceability~(cf.~\autoref{sec:related_work}).
Each dimension was derived through a transparent and systematic process and is illustrated with real-world examples from the cultural heritage.

Finally, we evaluated the practical applicability of the proposed DQ definition.
The results confirm that the defined dimensions capture the full spectrum of empirically observed quality problems.
All dimensions were represented in the catalog of quality problems, except for \emph{Relevance}, \emph{Confidentiality}, and \emph{Integrity}.
The absence of \emph{Relevance} suggests that domain experts generally assume maximal completeness of their datasets, considering all included information to be inherently relevant.
The omission of \emph{Confidentiality} likely reflects the field's strong open access and open science orientation, in which data is intended for unrestricted public use.
Similarly, \emph{Integrity} is of minor concern for object description data, as the contributing experts and organizations are primarily research-oriented and rarely encounter data corruption or tampering.
Although these three dimensions were not directly represented in the evaluated catalog, they retain both theoretical and practical relevance and could be further validated in additional assessments.
The largely orthogonal nature of the dimensions further supports the robustness and conceptual soundness of the proposed DQ definition.
Overall, this set of dimensions provides a comprehensive definition of DQ for object description data demonstrates and establishes a rigorous and practical foundation for systematic quality assessment within the cultural heritage domain.

This work is not necessarily limited to object description data.
While the proposed quality dimensions are specifically tailored to object description data, characterized by semi-structuredness, link quality, and domain-specific aspects such as provenance and traceability, these characteristics are not unique to the cultural heritage domain.
Similar features are found in other domains, such as biodiversity, which also manage heterogeneous, interlinked, and descriptive data.
Consequently, we anticipate that the proposed dimensions are applicable both within and beyond cultural heritage, providing a reference point for DQ assessment in related fields.

In order for organizations and users to evaluate their data against a DQ definition, quantitative measurability is essential.
A substantial body of literature already defines and applies metrics to assess the degree of quality across various dimensions.
Future research should clarify the extent to which existing metrics can be used to assess the defined quality dimensions, and identify aspects where new metrics are required.
For example, \citet{ehrlinger2022} provide a recent overview of tools for measuring and monitoring DQ, which could inform efforts to operationalize the dimensions of our definition.
Pattern-based approaches to quality problem detection, such as that of \citet{kesper2023}, may prove particularly effective for identifying structural quality problems, including heterogeneous representations of identical information.

Given the centrality of data in contemporary research and society, ongoing investigation into DQ remains essential.
While considerable research has already been conducted, the next step is to synthesize these efforts to develop a comprehensive quality assurance concept for cultural heritage, adaptable to the specific needs of diverse research fields and institutions.

%% file: appendix.tex
\appendix

\section{Classification of Data Quality Problems by Quality Dimension}

Table~\ref{tab:dataQualityProblems} provides additional details on the empirical evaluation presented in Section~\ref{sec:empirical_evaluation}.
It lists all data quality problems collected from domain experts, along with their unique identifier and the affected primary quality dimension.

{\scriptsize
  \setlength{\tabcolsep}{4pt}
  \begin{longtable}{>{\raggedright\arraybackslash}p{0.58\textwidth} >{\raggedright\arraybackslash}p{0.32\textwidth}}
    \caption{List of Data Quality Problems: ID, title, and primary quality dimension affected}\label{tab:dataQualityProblems} \\
    \toprule
    \textbf{Data Quality Problem}                                                     & \textbf{Primary Quality Dimension}    \\
    \midrule
    \endfirsthead

    \toprule
    \textbf{Data Quality Problem}                                                     & \textbf{Primary Quality Dimension}    \\
    \midrule
    \endhead

    \bottomrule
    \endfoot

    D01.1.1 Lack of data — empty fields                                               & Internal Completeness                 \\
    D01.1.2 Lack of data — incomplete fields                                          & Internal Completeness                 \\
    D01.1.3 Lack of records                                                           & Internal Completeness                 \\
    D01.1.4 Lack of source                                                            & Provenance                            \\
    D01.1.5 Lack of information about the person responsible for uncertain statements & Trustworthiness                       \\
    D01.1.6 Lack of metadata                                                          & Internal Completeness                 \\
    D01.1.7 Lack of rights statement and/or license                                   & Internal Completeness                 \\
    D01.1.8 No rating of a source in the data                                         & Trustworthiness                       \\
    D01.2 Unmarked multilingualism                                                    & Appropriateness                       \\
    D01.3.1 Heterogeneous structural representations                                  & Coherence                             \\
    D01.3.2 Heterogeneous precision of data                                           & Coherence                             \\
    D01.3.3 Heterogeneous qualifiers of uncertainty                                   & Coherence                             \\
    D01.3.4 Heterogeneous value representations                                       & Coherence                             \\
    D02.1 Spelling errors                                                             & Compliance                            \\
    D02.2 Incorrect information                                                       & Semantic Accuracy                     \\
    D02.3 Incorrect use of controlled vocabulary / authority file                     & External Accuracy                     \\
    D02.4 Incorrectly placed information                                              & Compliance                            \\
    D02.4.1 Multiple units of information in a single repeatable field                & Compliance                            \\
    D02.5.1 Non-matching date dependencies                                            & Logical Consistency                   \\
    D02.5.2 Non-matching functional dependencies of categorizations                   & Logical Consistency                   \\
    D02.5.3 Non-matching dependencies of spatial statements                           & Logical Consistency                   \\
    D02.5.4 Violation of dependencies between mandatory statements                    & Internal Completeness                 \\
    D03.1 Multiple data elements describing the same entity                           & Compactness                           \\
    D03.2 Redundancies in data                                                        & Compactness                           \\
    D04.1 Inconsistent use of units or metric systems                                 & Coherence                             \\
    D04.2 Missing units of measurement                                                & Internal Completeness                 \\
    D05.1 Missing references between data records                                     & External Completeness                 \\
    D05.2 Reference to a non-existent data record                                     & External Completeness                 \\
    D05.3 Ambiguous reference to a data record                                        & Logical Consistency                   \\
    D05.4 Ambiguous reference to described (real) entity                              & External Accuracy                     \\
    D05.5 Untraceable resource from URI namespaces                                    & Availability                          \\
    D06.1 Questionable data                                                           & Trustworthiness                       \\
    D06.2 Imprecision                                                                 & Precision                             \\
    D06.3 Contradiction                                                               & Logical Consistency                   \\
    D06.4 Unmarked uncertainties in data                                              & Internal Completeness                 \\
    D06.5 Implicitly marked uncertainties                                             & Compliance                            \\
    D06.6 Undescribed dependencies between uncertain statements                       & Semantic Accuracy                     \\
    D06.7 Lack of qualification of uncertainty                                        & Compliance                            \\
    D06.7.1 Indeterminate degree of uncertainty                                       & Compliance                            \\
    D06.8 Heterogeneous representations of uncertainty                                & Coherence                             \\
    D07.1 Undocumented data dynamics (in the data itself)                             & Temporal Traceability                 \\
    D07.2 Undocumented data changes caused by model dynamics                          & Temporal Traceability                 \\
    D07.3 Outdated data                                                               & Data Currency                         \\
    D08 Subjectivity                                                                  & Trustworthiness                       \\
    D09 Implicit knowledge                                                            & Internal Completeness                 \\
    D09.1 Use of non-standard symbols to express certain facts                        & Appropriateness                       \\
    D10.1 Violation of controlled vocabularies (use of custom values)                 & Syntactic Accuracy                    \\
    D10.2 Lack of reference to authority data (global comparability)                  & Compliance                            \\
    D10.3 Unnecessary use of custom controlled vocabulary                             & Compliance                            \\
    D11 Violation of format specifications                                            & Compliance                            \\
    D12 Incompatible data types                                                       & Syntactic Accuracy                    \\
  \end{longtable}
}

\section{Chi-Square Values for Pairwise Independence of Data Quality Dimensions}

Table~\ref{tab:chi2DataQualityDimensions} shows the chi-square values calculated for each pair of data quality dimensions during the evaluation of their orthogonality (cf.~Section~\ref{sec:empirical_evaluation}).

\begin{landscape}
  \begin{table}
    \renewcommand{\arraystretch}{1.5}
    \caption{Chi-square values for each pair of data quality dimensions.
      Only the lower triangle is shown, since the matrix is symmetrical.
      Dimensions with an occurrence equal to zero were excluded, since they are irrelevant for the Chi-square test.}\label{tab:chi2DataQualityDimensions}
    \scriptsize
    %\begin{tabular}{lrr}
    %\begin{table}[!ht]
    %\centering
    \begin{tabular}{lllllllllllllllllllll}
      \toprule
                                                      &
      \rotatebox{90}{\textbf{syntactic accuracy}}     &
      \rotatebox{90}{\textbf{semantic accuracy}}      &
      \rotatebox{90}{\textbf{external accuracy}}      &
      \rotatebox{90}{\textbf{compliance}}             &
      \rotatebox{90}{\textbf{precision}}              &
      \rotatebox{90}{\textbf{internal completeness}}  &
      \rotatebox{90}{\textbf{external completeness}}  &
      \rotatebox{90}{\textbf{compactness}}            &
      \rotatebox{90}{\textbf{logical consistency}}    &
      \rotatebox{90}{\textbf{coherence}}              &
      \rotatebox{90}{\textbf{availability}}           &
      \rotatebox{90}{\textbf{data currency}}          &
      \rotatebox{90}{\textbf{data update currency}}   &
      \rotatebox{90}{\textbf{time concurrency}}       &
      \rotatebox{90}{\textbf{provenance}}             &
      \rotatebox{90}{\textbf{trustworthiness}}        &
      \rotatebox{90}{\textbf{temporal traceability}}  &
      \rotatebox{90}{\textbf{causal traceability}}    &
      \rotatebox{90}{\textbf{appropriateness}}        &
      \rotatebox{90}{\textbf{versatility}}                                                                                                                                                \\
      \midrule
      syntactic accuracy                             & ~    & ~    & ~    & ~    & ~    & ~    & ~    & ~    & ~    & ~    & ~    & ~    & ~    & ~    & ~    & ~    & ~    & ~    & ~    \\
      semantic accuracy                              & 0,03 & ~    & ~    & ~    & ~    & ~    & ~    & ~    & ~    & ~    & ~    & ~    & ~    & ~    & ~    & ~    & ~    & ~    & ~    \\
      external accuracy                              & 0,64 & 0,02 & ~    & ~    & ~    & ~    & ~    & ~    & ~    & ~    & ~    & ~    & ~    & ~    & ~    & ~    & ~    & ~    & ~    \\
      compliance                                     & 2,00 & 0,48 & 0,50 & ~    & ~    & ~    & ~    & ~    & ~    & ~    & ~    & ~    & ~    & ~    & ~    & ~    & ~    & ~    & ~    \\
      precision                                      & 0,01 & 0,90 & 1,80 & 2,60 & ~    & ~    & ~    & ~    & ~    & ~    & ~    & ~    & ~    & ~    & ~    & ~    & ~    & ~    & ~    \\
      internal completeness                          & 0,50 & 0,71 & 0,05 & 0,03 & 1,80 & ~    & ~    & ~    & ~    & ~    & ~    & ~    & ~    & ~    & ~    & ~    & ~    & ~    & ~    \\
      external completeness                          & 0,12 & 4,50 & 9,50 & 2,10 & 1,30 & 9,10 & ~    & ~    & ~    & ~    & ~    & ~    & ~    & ~    & ~    & ~    & ~    & ~    & ~    \\
      compactness                                    & 0,01 & 2,40 & 0,28 & 1,00 & 0,81 & 0,14 & 0,89 & ~    & ~    & ~    & ~    & ~    & ~    & ~    & ~    & ~    & ~    & ~    & ~    \\
      logical consistency                            & 0,18 & 2,90 & 0,01 & 2,80 & 1,30 & 0,17 & 0,53 & 0,73 & ~    & ~    & ~    & ~    & ~    & ~    & ~    & ~    & ~    & ~    & ~    \\
      coherence                                      & 0,02 & 5,00 & 0,01 & 4,20 & 0,64 & 3,90 & 0,15 & 0,01 & 2,10 & ~    & ~    & ~    & ~    & ~    & ~    & ~    & ~    & ~    & ~    \\
      availability                                   & 0,17 & 1,70 & 6,30 & 0,31 & 0,59 & 0,86 & 7,70 & 2,90 & 1,40 & 0,98 & ~    & ~    & ~    & ~    & ~    & ~    & ~    & ~    & ~    \\
      data currency                                  & 0,98 & 1,40 & 0,75 & 0,40 & 0,59 & 0,08 & 2,10 & 2,90 & 0,43 & 0,17 & 0,20 & ~    & ~    & ~    & ~    & ~    & ~    & ~    & ~    \\
      data update currency                           & 0,31 & 1,90 & 0,19 & 0,98 & 0,19 & 0,71 & 0,43 & 0,09 & 0,14 & 0,31 & 0,06 & 16,3 & ~    & ~    & ~    & ~    & ~    & ~    & ~    \\
      time concurrency                               & 0,98 & 0,01 & 0,59 & 0,31 & 0,59 & 0,86 & 0,02 & 2,90 & 9,30 & 0,17 & 0,20 & 4,30 & 0,06 & ~    & ~    & ~    & ~    & ~    & ~    \\
      provenance                                     & 1,30 & 0,20 & 0,28 & 4,20 & 0,81 & 0,14 & 0,89 & 0,37 & 0,58 & 0,01 & 0,27 & 0,27 & 0,09 & 0,27 & ~    & ~    & ~    & ~    & ~    \\
      trustworthiness                                & 2,90 & 0,02 & 0,62 & 9,10 & 1,80 & 0,30 & 1,90 & 0,81 & 0,01 & 0,64 & 0,59 & 0,59 & 0,19 & 0,59 & 23,3 & ~    & ~    & ~    & ~    \\
      temporal traceability                          & 0,98 & 1,70 & 0,59 & 3,10 & 0,59 & 0,86 & 0,02 & 0,27 & 1,40 & 0,98 & 0,20 & 0,20 & 0,06 & 4,30 & 2,90 & 0,75 & ~    & ~    & ~    \\
      causal traceability                            & 0,64 & 1,10 & 0,39 & 2,00 & 0,39 & 3,00 & 0,43 & 0,18 & 2,90 & 0,64 & 0,13 & 0,13 & 0,04 & 7,30 & 5,10 & 1,90 & 33,3 & ~    & ~    \\
      appropriateness                                & 4,90 & 0,08 & 0,35 & 8,60 & 0,91 & 1,20 & 0,43 & 0,85 & 6,00 & 4,90 & 0,24 & 0,49 & 1,10 & 0,24 & 0,85 & 4,50 & 0,24 & 0,01 & ~    \\
      versatility                                    & 4,10 & 0,10 & 1,03 & 2,13 & 0,08 & 0,00 & 2,30 & 0,47 & 0,74 & 0,84 & 0,35 & 0,35 & 0,11 & 0,35 & 0,47 & 1,03 & 0,35 & 0,23 & 2,41 \\
      \bottomrule
    \end{tabular}
  \end{table}
\end{landscape}